\documentclass[preprint,floatfix,showpacs,showkeys]{revtex4-1}

\usepackage{graphicx}
\usepackage{xcolor}
\usepackage{amsmath}
\usepackage{amssymb}

\begin{document}

\title{Kink-antikink soliton solutions of the nonlinear Klein-Gordon equation on branched structures}
\author{$^{1}$Q.U.~Asadov, $^{2,*}$K.K.~Sabirov, $^{3}$J.R.~Yusupov}
\affiliation{$^1$ TMS Institute Tashkent, 41-b Mukimi street, Tashkent, Uzbekistan\\
$^2$University of Tashkent for Applied Sciences,
1- Gavkhar street, 100149 Tashkent, Uzbekistan\\
$^{3}$Kimyo International University in Tashkent, Shota Rustaveli str. 156, Tashkent 100121, Uzbekistan}

\email{Corresponding author: karimjonsabirov80@gmail.com}

 % insert PACS

\begin{abstract}
In this paper, we investigate the nonlinear Klein–Gordon equation on a metric star graph with three semi-infinite bonds. At the branching point, we impose  a weighted continuity condition and a generalized weighted Kirchhoff condition for the derivatives of the wave function. By employing both analytical methods and numerical techniques, we construct exact and numerical soliton solutions that satisfy the vertex conditions and conserve energy and momentum. The results of analytic calculations are confirmed through numerical experiments, which demonstrate reflectionless propagation of kink–antikink soliton solutions. We compute and analyze the reflection coefficient, study the impact of various nonlinearity parameters, and further extend the formulation to other graph topologies, such as tree and loop graphs.
\end{abstract}


\maketitle

\section{Introduction}

The nonlinear dynamics of solitary waves, which are commonly described by nonlinear partial differential equations (PDEs), appear in many branches of science and engineering,  such as fluid dynamics, optics, plasma physics, and field theory, where they manifest as localized, stable structures that propagate without changing shape. A classical mathematical framework for understanding these phenomena is provided by integrable systems theory and the inverse scattering transform, as developed in the pioneering works \cite{Ablowitz1,Ablowitz2}.

Among nonlinear PDEs, the nonlinear Klein–Gordon (NLKG) equation represents a particularly important class. It arises naturally in the relativistic theory of scalar fields, where it models the behavior of elementary particles, and in solid-state physics, where it describes the propagation of dislocations in crystal lattices \cite{Greiner}. 
The equation has been studied from various perspectives, including its exact soliton-like solutions obtained under separation of variables \cite{Grundland}, its analytical treatment in higher-dimensional settings \cite{Matsuno}, and investigations into the integrability of specific forms of the equation \cite{Clarkson}.
One of the earliest numerical studies of the NLKG equation was conducted in \cite{Ablowitz3}, where authors examined solitary wave collisions in the context of nonlinear lattice equations, including discretized Klein–Gordon systems. Their scheme employed a Taylor series-based discretization with time and space steps $k = h = 0.05$, laying the foundation for numerical soliton simulations and collision studies. Further developments focused on the construction of numerical methods that preserve invariants of the continuous model, such as energy and momentum. In the Ref.~\cite{Jim} a comparative analysis of four different finite-difference schemes applied to the NLKG equation, emphasizing their ability to conserve energy over long integration times and assessing their stability and accuracy is provided. Invariant-preserving algorithms, formulating finite difference methods that exactly conserve discrete analogs of the NLKG's continuous invariants are proposed in \cite{Los}. These schemes have since been widely adopted by researchers and are commonly used as benchmark methods in the study of NLKG equation.
 
Despite the extensive study of the nonlinear Klein–Gordon equation from various analytical and numerical perspectives, its formulation and analysis on branched structures remains relatively unexplored. 
The nonlinear PDEs on the branched
structures have attracted the most interest in the last decades. This attention was caused by the possibility to obtain the soliton solution of nonlinear partial differential equations such
as nonlinear Schr\"odinger, Dirac equations and its numerous
applications in different branches of physics
\cite{Sobirov1,Adami,Sobirov2,Sabirov1,Sabirov2,Aripov2019,Yusupov2019}. The branched
structures can be modeled by metric graphs. Metric graphs consist
of two sets: the set of $V$ vertices (points) and the set of $B$ bonds (intervals)
that connect pairs of vertices. The topology of the graphs can be determined by the so-called
adjacency matrix $C=(C_{ij})$ \cite{Kottos,Gnutzmann,Berkolaiko}:
$$
C_{ij}=\left\{\begin{array}{ll}1,\, \textrm{if } i \textrm{ and } j\, \textrm{are connected,}\\0,\,\textrm{otherwise,}\end{array}\right. i,j=1,2,...,V.
$$

Recent progress in the analysis of wave phenomena on metric graphs has led to several new developments, including results on the stability of standing waves, the existence of breather solutions, and the characterization of small-amplitude oscillations on such networks \cite{Maier,Kumbi1,Takei,Golosh1,Alrazi}. In this paper, we focus on one of the exact solutions of the nonlinear Klein-Gordon equation with cubic nonlinearity and study its transmission through a vertex of a graph. 

The paper is organized as follows. In the next section, we recall the nonlinear Klein–Gordon equation and its kink and soliton solutions, with a focus on the case of cubic nonlinearity. Section 3 presents the formulation of the nonlinear Klein–Gordon equation on a metric star graph, including the derivation of appropriate vertex boundary conditions. In Section 4, we analytically construct a kink soliton solution to the problem and establish a sum rule that ensures reflectionless transmission. The obtained results are confirmed with numerical experiments. Section 5 is devoted to extending the analysis to other graph topologies, such as tree and loop graphs. Finally, the last section provides concluding remarks.

\section{The nonlinear Klein-Gordon equation}

The nonlinear Klein-Gordon equation has attracted significant interest in  mathematical physics, serving as a fundamental model for describing scalar field dynamics in various physical contexts, including quantum field theory, condensed matter physics, and cosmology. This PDE extends the linear Klein-Gordon equation by including nonlinear terms, which give rise to complex phenomena such as solitons, wave interactions, and energy localization. In its general form, the equation is expressed as
\begin{equation}
    \partial^2_{t}u-\partial^2_{x}u+F'(u)=0
\end{equation}
where $\partial_t^2=\partial^2/\partial t^2$, $\partial_x^2=\partial^2/\partial x^2$ and $F'(u)=dF(u)/du$ (derivative of a potential energy $F(u)$) is a nonlinear function of $u = u(t, x)$. The nonlinear term $F'(u)$ introduces rich dynamical behaviors, making the equation a subject of extensive study in both theoretical and computational frameworks. For example, a choice $F'(u)=\sin{(u)}$ gives the famous sine-Gordon equation.
Numerical simulations of the equation with various nonlinear potentials, including the sine-Gordon and cubic forms have been carried out in \cite{Ablowitz3}. 

Among the various nonlinearities, the cubic nonlinearity is of particular interest due to its well-established theoretical foundation. In this context, the NLKG equation with cubic nonlinearity can be written as \cite{Wazwaz}
\begin{equation}\label{eq::cnkge}
\partial^2_{t} u - a^2 \partial^2_{x} u + a u - b u^3 = 0,
\end{equation}
where $a, b \in \mathbb{R}$, $ab \ne 0$. The nonlinearity here stems from the potential function $F(u) = \frac{a}{2}u^2 - \frac{b}{4}u^4$, which exhibits a double-well structure when $a, b > 0$.

The equation admits two principal types of traveling wave solutions derived via the ansatz $u(x,t) = U(\xi)$ with $\xi = \mu(x - ct)$, where $c$ is the wave velocity and $\mu$ the wave number. Substitution yields an ordinary differential equation for $U(\xi)$, from which kink and soliton solutions emerge, depending on the sign and magnitude of the parameters.

The kink solution is obtained using the tanh method \cite{Wazwaz} and given by
\begin{equation}\label{eq::kink}
u(x, t) = \sqrt{\frac{a}{b}} \tanh\left[\sqrt{\frac{a}{2(c^2 - a^2)}} (x - ct - x_0)\right],
\end{equation}
where $c^2 - a^2>0$, with $x_0$ denoting the initial wave position.

On the other hand, the soliton solution, which is found by applying sech method \cite{Wazwaz} takes the form
\begin{equation}\label{eq::soliton}
u(x, t) = \sqrt{\frac{2a}{b}} \text{sech}\left[\sqrt{\frac{a}{a^2 - c^2}} (x - ct - x_0)\right],
\end{equation}
where $a^2 - c^2>0$.

An important feature of this system is the conservation of energy and momentum. The energy for the equation \eqref{eq::cnkge} (for $x\in D \subset \mathbb{R}$) is expressed as
\begin{equation}\label{eq::energy}
E(t) = \frac{1}{2} \int_{D} \left[ (\partial_t u)^2 + a^2 (\partial_x u)^2 + a u^2 - \frac{b}{2} u^4 \right] dx,
\end{equation}
and the linear momentum is
\begin{equation}\label{eq::moment}
P(t) = \int\limits_{-\infty}^{+\infty}\partial_t u\  \partial_x u\ dx,
\end{equation}
which remain invariant for all $t > 0$ under appropriate boundary conditions. This conserved quantities are critical measures for both theoretical analysis and numerical accuracy.

\section{The nonlinear Klein-Gordon equation on a star graph}

\begin{figure}[t!]
\centering
\includegraphics[width=8cm]{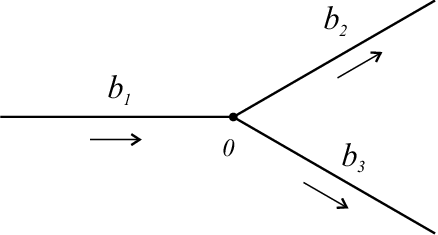}
\caption{The metric star graph} \label{pic1}
\end{figure}

We consider a star graph shown in Fig.~\ref{pic1} with three bonds $b_j$ ($j=1,2,3$), for which coordinates $x_j$ are assigned. Choosing the origin of coordinates
at the vertex, 0, for the bond $b_1$ we put $x_1\in(-\infty,0]$ and
for $b_{2,3}$ we fix $x_{2,3}\in[0,+\infty)$. In what follows, we
use the shorthand notation $u_{j}(x)$ for $u_j(x_j)$ where $x$ is
the coordinate on the bond $j$ to which the component $u_j$
refers. The NLKG equation on each bond $b_j$ of the star
graph is written as
\begin{equation}
\partial_{t}^2 u_j - \partial_{x}^2 u_j - u_j + \beta_{j} u_{j}^3 =
0,\,\beta_j>0.\label{kgeq1}
\end{equation}

To solve the problem one needs to define the vertex boundary conditions at the branched point
of the star graph. To this end, we derive physically acceptable boundary conditions
from conservation laws, such as energy conservation, which is defined as
\begin{equation}
E = \sum_{j=1}^3 E_j,\label{conl1}
\end{equation}
where $E_j$ are partial energies defined by Eq.~\eqref{eq::energy} as
\begin{equation}
E_j=\frac{1}{2} \int _{b_j} \left[  (\partial_t u_j)^2 +
(\partial_x u_j)^2 - u_j^2 + \frac{\beta_j}{2}
u_j^4 \right] dx.\label{energy1}
\end{equation}

The energy conservation implies $\dot{E}=0$, which results in nonlinear boundary
conditions given as
\begin{equation}
\partial_x u_1 \partial_t u_1 \big|_{x=0} = \partial_x u_2 \partial_t u_2 \big|_{x=0} + \partial_x u_3 \partial_t u_3 \big|_{x=0}.\label{nbc1}
\end{equation}
To solve the Eq.~\eqref{kgeq1} it is required two boundary conditions and at the same time the nonlinear vertex boundary
condition \eqref{nbc1} must be fulfilled. The two boundary conditions that met this requirement is the weighted continuity
\begin{equation}
 \alpha_1 u_1 \big|_{x=0}= \alpha_2 u_2 \big|_{x=0} = \alpha_3 u_3 \big|_{x=0},\label{wc1}
 \end{equation}
and weighted current conservation conditions (considered also in \cite{Berkolaiko})
\begin{equation}
\frac{1}{\alpha_1} \partial_x u_1 \big|_{x=0} = \frac{1}{\alpha_2} \partial_x u_2 \big|_{x=0} + \frac{1}{\alpha_3} \partial_x u_3 \big|_{x=0},\label{kr1}
\end{equation}
with $\alpha_j\not=0,\,j=1,2,3$ being either all positive or all negative real constants. The boundary conditions \eqref{wc1} and \eqref{kr1} comply with the condition \eqref{nbc1}, the proof of which is provided in Appendix A. 

Let us suppose that there exists a universal function 
$g(x, t)$, independent of the bond, and serves as a common solution for the star graph, which satisfies 
\begin{align}
\alpha_j u_j(0,t)&=g(0,t),\nonumber\\
\alpha_j \partial_x u_j(x,t) \big|_{x=0}&=\partial_x g(x,t) \big|_{x=0},\label{bcgf}
\end{align}
for $j=1,2,3$.
Using these relations one can rewrite Eq.~\eqref{kr1} as
$$
\frac{1}{\alpha_1} \frac{1}{\alpha_1} \partial_x g(x,t) \big|_{x=0} = \frac{1}{\alpha_2} \frac{1}{\alpha_2} \partial_x g(x,t) \big|_{x=0} + \frac{1}{\alpha_3} \frac{1}{\alpha_3} \partial_x g(x,t) \big|_{x=0},
$$
and what follows from this is 
$$
\frac{1}{\alpha_1 ^2} = \frac{1}{\alpha_2 ^2} + \frac{1}{\alpha_3 ^2}.
$$
In particular, one can introduce the solution of Eq.~\eqref{kgeq1} given on the star graph, in the form
\begin{equation}\label{eq::gensol}
u_j (x,t) = \frac{1}{\sqrt{\beta_j}} \phi (x,t),
\end{equation}
where $\phi$ is $\beta$-independent universal solution, which satisfies the nonlinear Klein-Gordon equation
$$
\partial_{t}^2 \phi - \partial_{x}^2 \phi - \phi + \phi^3 = 0,\ \ -\infty<x<+\infty.
$$
Here, the functions $\phi(x_1,t)$ and $\phi(x_{2,3},t)$ are defined on $(-\infty,0]$ and $[0,+\infty)$, respectively.
Next, choosing the boundary conditions parameters as
$$
\alpha _j = \sqrt{\beta_j} , \,\,\, (j=1,2,3),
$$
one can rewrite Eqs.~\eqref{bcgf} as
\begin{align*}
\sqrt{\beta_j} u_j \big|_{x=0} &= \phi(0,t),\\
\sqrt{\beta_j} \partial_x u_j |_{x=0} &= \partial_x \phi (x,t)\big|_{x=0},
\end{align*}
which leads to the following sum rule for the nonlineary coefficients
\begin{equation}
    \frac {1}{\beta_1} = \frac {1}{\beta_2} + \frac {1}{\beta_3} . \label{sr1}
\end{equation}

Thus, the general solution \eqref{eq::gensol}, together with boundary conditions satisfying the constraint \eqref{sr1}, ensures energy conservation on the star graph.

It should be noted, that though the results obtained regards the star graph with one negative and two positive semi-infinite bonds, derivation of the sum rule (constraint in the form \eqref{sr1}) can be readily extended to a star graph with one negative and $N-1$ positive semi-infinite bonds. In this more general case, the sum rule takes the form:
$$
\frac {1}{\beta_1} = \sum\limits_{j=2}^{N}\frac {1}{\beta_j}.
$$

\section{Propagation of the kink soliton in a star graph}

Referring to Eq.~\eqref{eq::kink} in Section 2, the kink (antikink) soliton solution of NLKG equation
(\ref{kgeq1}) on each bond $b_j$ of a metric star graph can be written as
\begin{equation}
 u_j (x,t) = \mp \frac{1}{\sqrt{\beta_j}} \, {\rm tanh} \left( \frac{x - l - v t}
{\sqrt{2(1 - v^2) }} \right),\label{sol1}
\end{equation}
where $l$ is the center of the kink (antikink), $v\in(0;1)$ is its velocity. Fulfilling the vertex boundary conditions
(\ref{wc1})-(\ref{kr1}) leads to the following constrains
\begin{align}
&\frac {\alpha_1}{\sqrt{\beta_1}} = \frac {\alpha_2}{\sqrt{\beta_2}} =
\frac{\alpha_3}{\sqrt{\beta_3}},\label{const1}\\
&\frac {1}{\alpha_1 \sqrt{\beta_1}} = \frac {1}{\alpha_2 \sqrt{\beta_2}} +
\frac{1}{\alpha_3 \sqrt{\beta_3}},\label{const2}
\end{align}
where $l_j=l,\,v_j=v,\,j=1,2,3$.

From (\ref{const1}) and (\ref{const2}) for the reflectionless propagation of the kink soliton (\ref{sol1}) of the Klein-Gordon equation (\ref{kgeq1}) on the star graph we obtain the sum rule for nonlinearity coefficients as (\ref{sr1}).

\begin{figure}[t!]
\begin{minipage} [t!]{0.49\linewidth}
\centering
\includegraphics[width=1\linewidth]{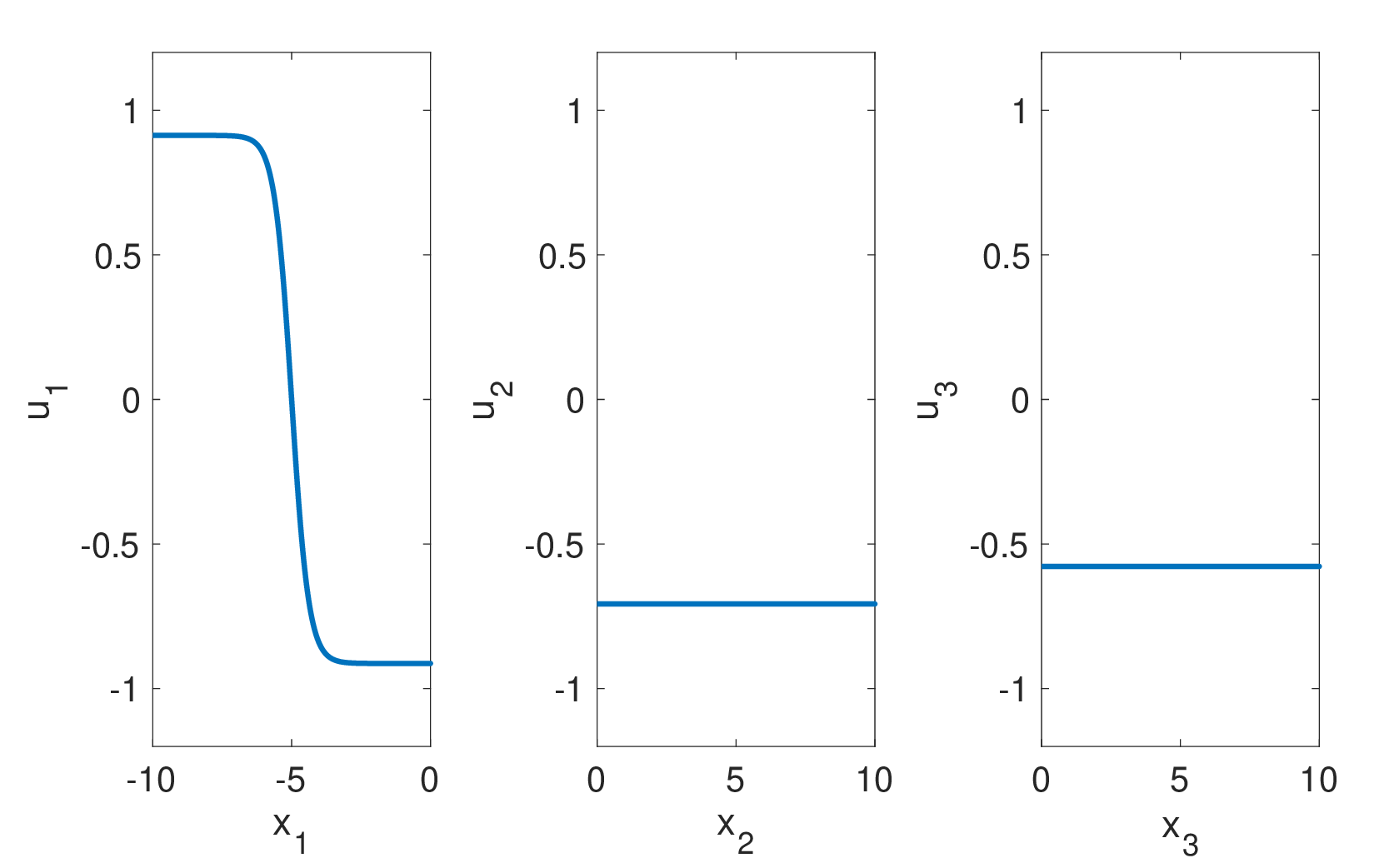} \\ a)
\end{minipage}
\hfill
\begin{minipage}[t!]{0.49\linewidth}
\centering
\includegraphics[width=1\linewidth]{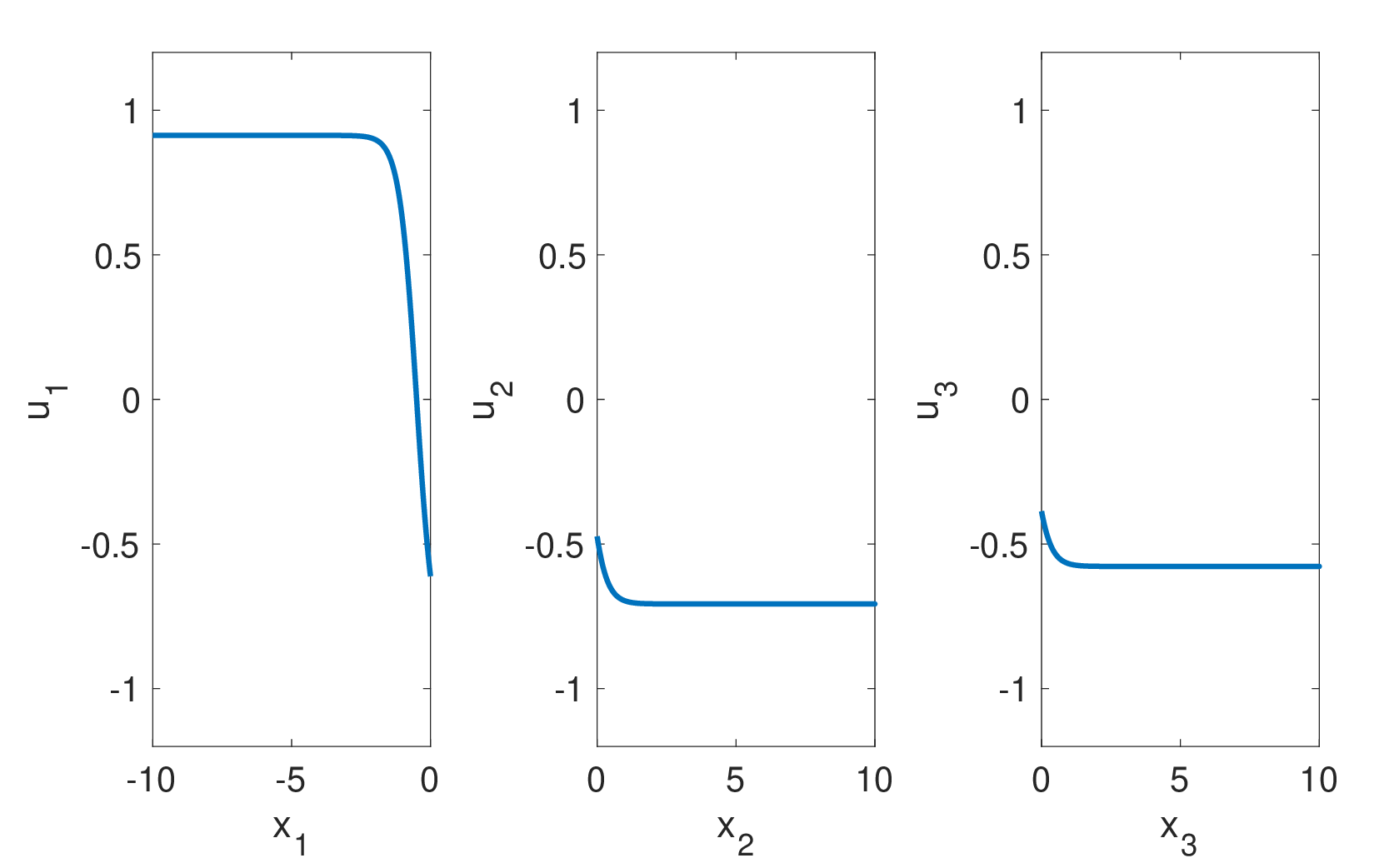} \\ b)
\end{minipage}
\vfill
\begin{minipage}[t!]{0.49\linewidth}
\centering
\includegraphics[width=1\linewidth]{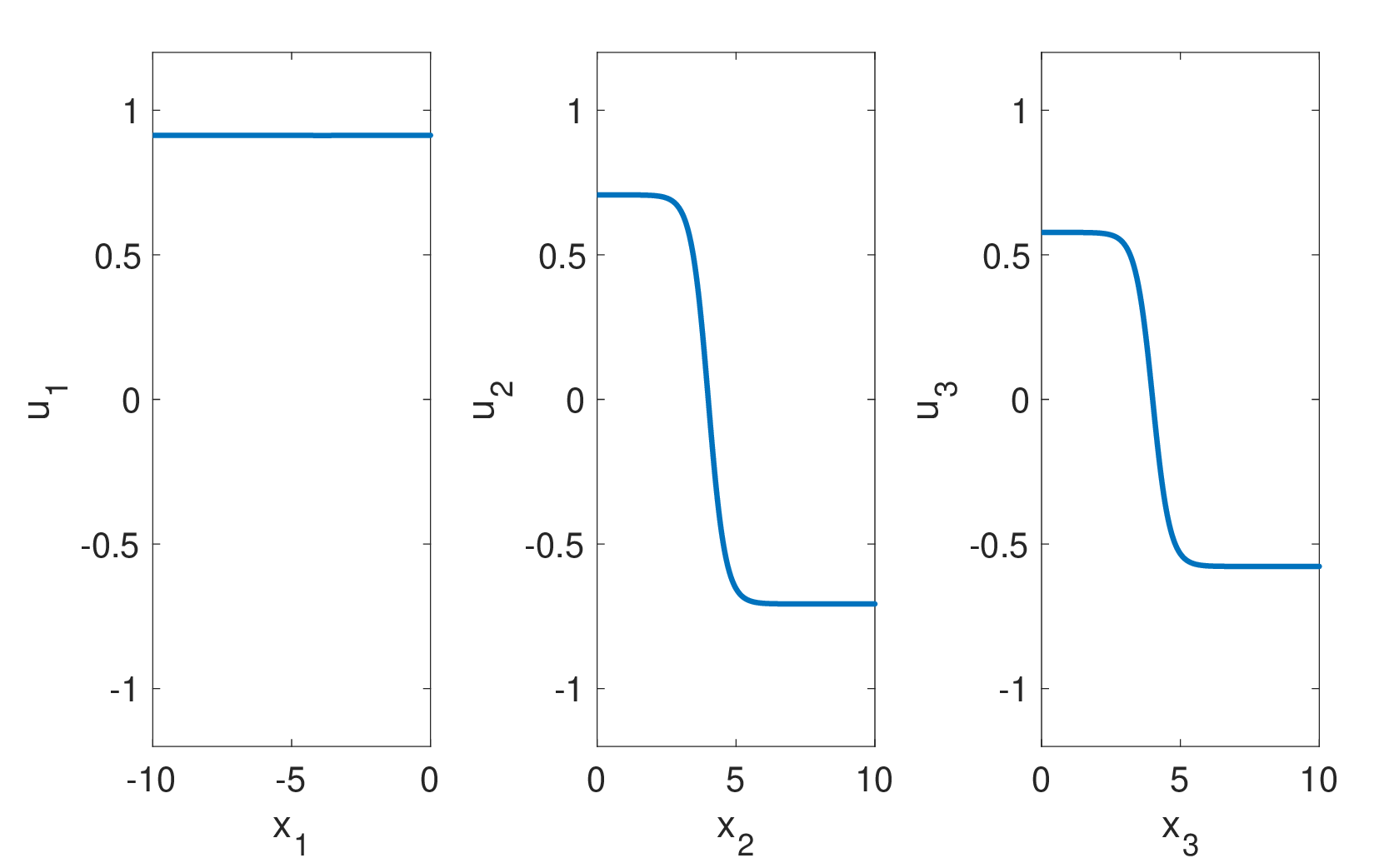} \\ c)
\end{minipage}

\caption{Propagation of the kink soliton, the center of which is initially located at $l=-5$ in the first bond with velocity $v=0.9$ for the three time moments a)~t=0; b)~t=5; c)~t=10. For the chosen nonlinearity coefficients, $\beta_1=1.2$, $\beta_2=2$, $\beta_3=3$ (the sum rule is satisfied), the kink propagates without reflection through the vertex.}
\label{pic2}
\end{figure}

\begin{figure}[h!]
\centering
\includegraphics[scale=0.6]{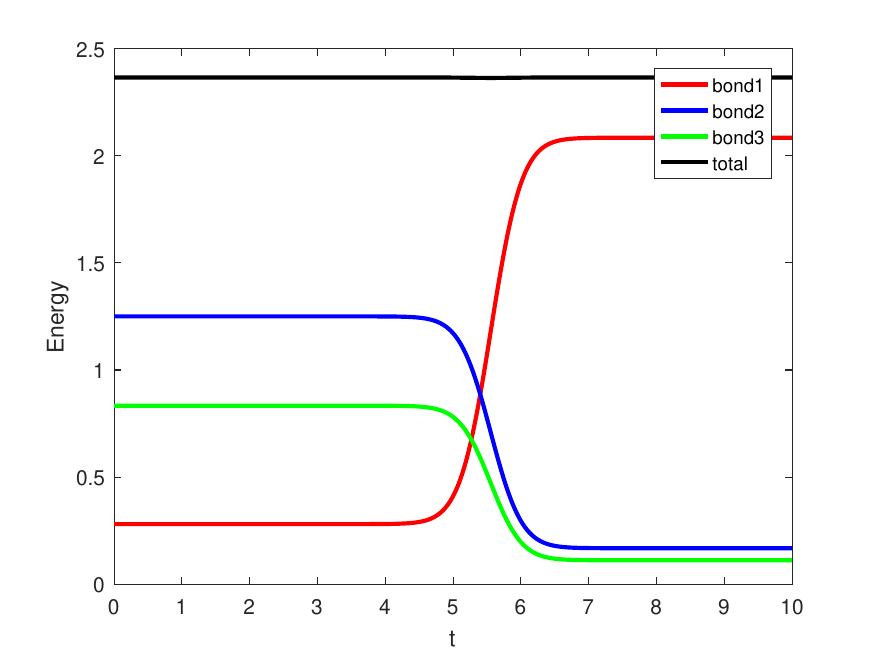}
\caption{(Color online) The energies on the each bonds and the total energy}
\label{pic3}
\end{figure}

Let us perform a numerical experiment by choosing kink soliton solution as initial condition and study its dynamics focusing on transition through the branching point (vertex). For this, we use a Taylor series-based discretization (see, Appendix B). Consider the initial set up with the kink centered at position $l=-5$ (in the first bond $b_1$) propagating with the velocity $v=0.9$. The nonlinearity parameters chosen to satisfy the sum rule constraint \eqref{sr1}, with values $\beta_1=1.2,\,\beta_2=2,\,\beta_3=3$. Figure~\ref{pic2} illustrates the kink dynamics up to time $t=10$, under the described setup, demonstrating that the kink passes through the vertex without any reflection. Accordingly, one can compute the time dependence of partial energies, the sum of which (being the total energy) is expected to be conserved over time. This is shown in Fig.~\ref{pic3}, which confirms the expected behavior.

\begin{figure}[t!]
\begin{minipage} [t!]{0.49\linewidth}
\centering
\includegraphics[width=1\linewidth]{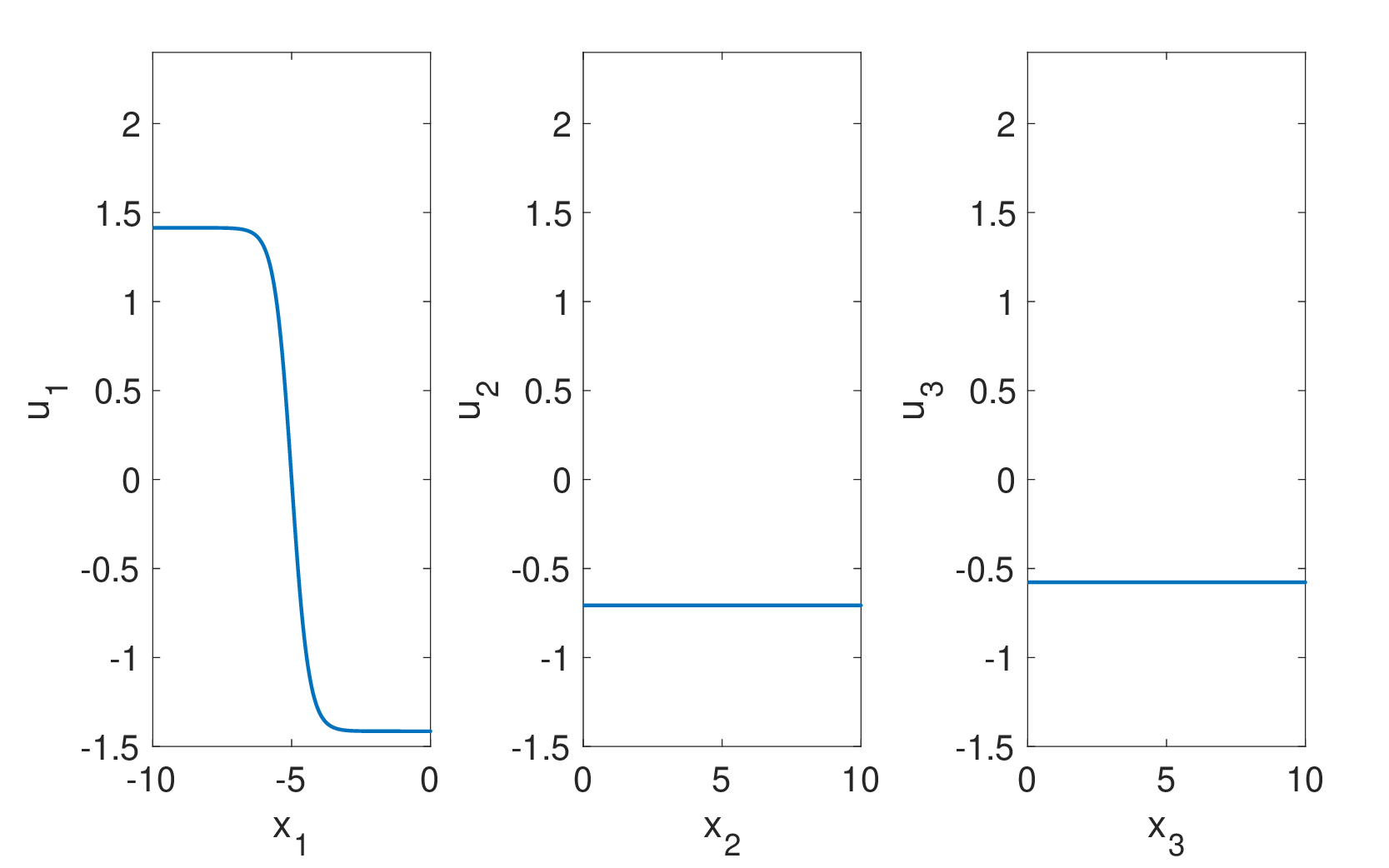} \\ a)
\end{minipage}
\hfill
\begin{minipage}[t!]{0.49\linewidth}
\centering
\includegraphics[width=1\linewidth]{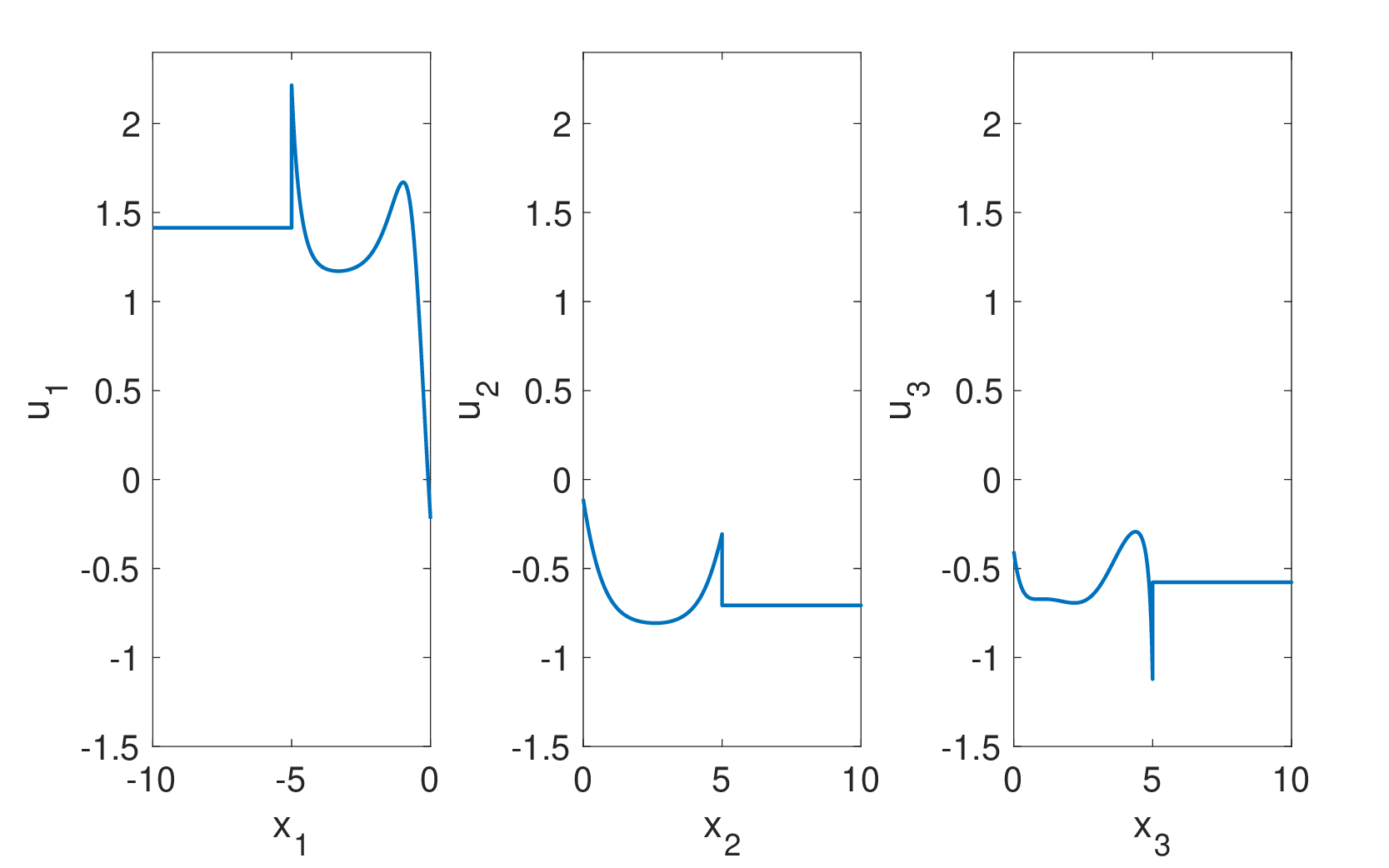} \\ b)
\end{minipage}
\vfill
\begin{minipage}[t!]{0.49\linewidth}
\centering
\includegraphics[width=1\linewidth]{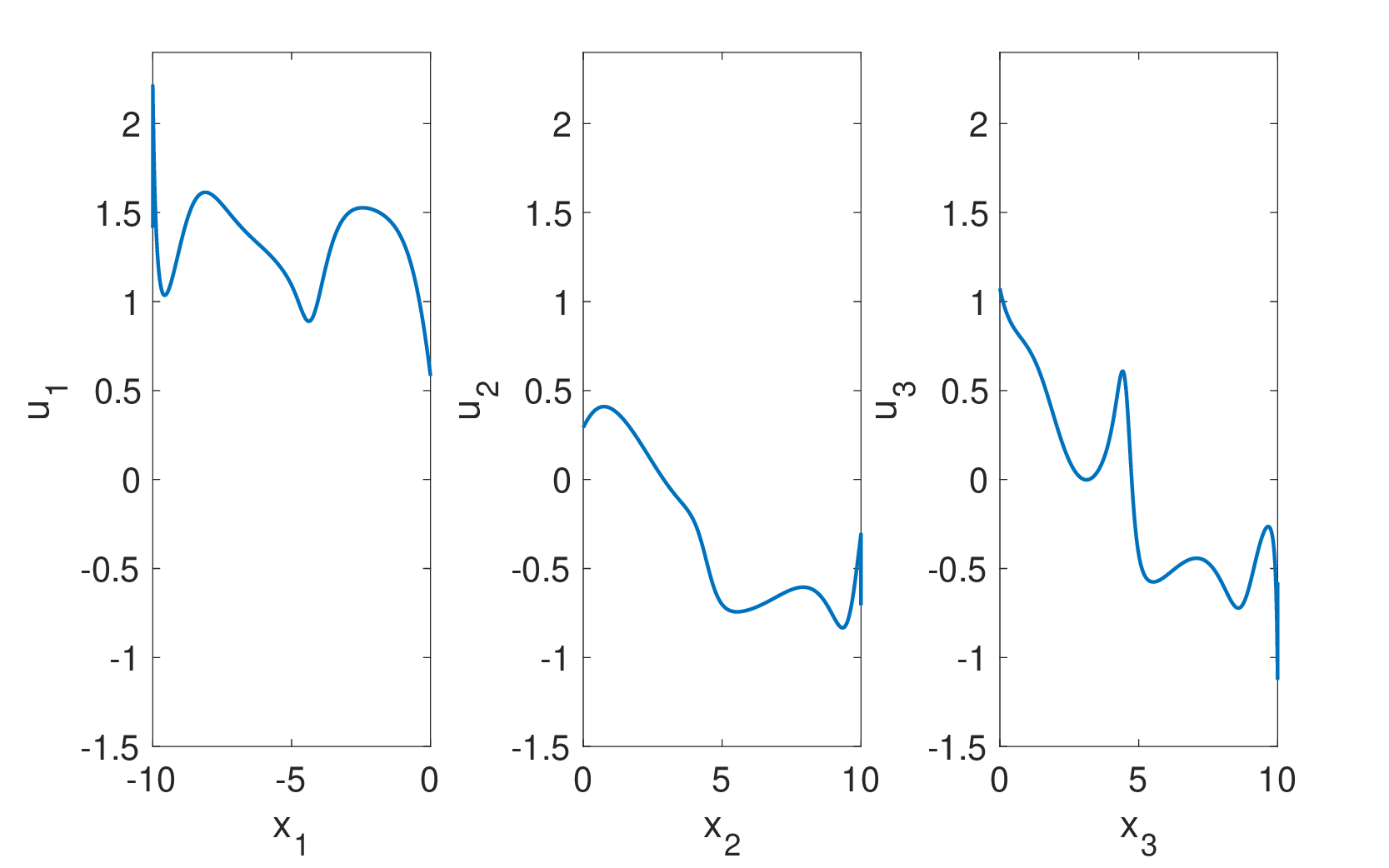} \\ c)
\end{minipage}

\caption{Propagation of the kink with the same initial conditions as in Fig.~\ref{pic2}, but for the nonlinearity coefficients $\beta_1=0.5$, $\beta_2=2$, $\beta_3=3$ (the sum rule is not satisfied). The kink propagation is not reflectionless anymore.}
\label{pic4}
\end{figure}

Now, it remains to demonstrate that when the nonlinearity coefficients do not satisfy the sum rule, the kink dynamics exhibit reflection at the branching point. To this end, we numerically solve the NLKG equation \eqref{kgeq1} using parameters that violate the sum rule — specifically, we choose $\beta_1 = 0.5$, $\beta_2 = 2$, and $\beta_3 = 3$. Figure~\ref{pic4} illustrates the propagation of the kink with the same initial conditions as in Fig.~\ref{pic2}, but for the above set of nonlinearity coefficients. It is evident from the figure that the kink propagation is no longer reflectionless.

While the plot in Fig.~\ref{pic2} above shows that the kink soliton propagates without reflection, this observation can be further verified quantitatively by calculating the reflection coefficient on the first bond. For this we turn to the time dependence of the linear momentum defined in \eqref{eq::moment}. For the considered star graph the total momentum can be derived analytically: 
\begin{align}
&P_\text{tot}(t) = \sum_{j=1}^3P_j(t) =\sum_{j=1}^3 \int_{b_j} \partial _t u_j \partial _x u_j dx \\
&=-\frac{\upsilon }{\sqrt{2(1-\upsilon^2)}} \left[\frac{2}{3}\left(\frac{1}{\beta_1}+\frac{1}{\beta_2} + \frac{1}{\beta_3}\right)+\left(\frac{1}{\beta_1}-\frac{1}{\beta_2} - \frac{1}{\beta_3}\right) \left(\tanh \left(g(t) \right) - \frac{1}{3} \tanh ^3 \left(g(t)\right) \right)\right],\nonumber
\end{align}
where $P_j$ are partial momentum related to each bond and
$$
g(t) = \frac{-l-\upsilon t}{\sqrt{2(1-\upsilon ^2)}}.
$$

It is easy to see, that when the sum rule \eqref{sr1} for the nonlinearity coefficients of the system is satisfied, the total momentum remains constant (time-independent), and is therefore conserved:
\begin{equation*}
P_\text{tot}(t)\Big|_{\frac {1}{\beta_1} = \frac {1}{\beta_2} + \frac {1}{\beta_3}} = -\frac{4 \upsilon }{ 3\beta_1 \sqrt{2(1-\upsilon^2)} }.
\end{equation*}

The time dependence of the reflection coefficient can be defined as
\begin{align}\label{RCmomentum}
R(t)&=\frac{P_1(t)}{P_\text{tot}(t=0)}\nonumber\\
&=\frac{ \frac{1}{\beta_1} \left(\tanh \left(g(t)\right) - \frac{1}{3} \tanh ^3 \left(g(t)\right) + \frac{2}{3} \right)}{ \frac{2}{3}\left(\frac{1}{\beta_1}+\frac{1}{\beta_2} + \frac{1}{\beta_3}\right)+\left(\frac{1}{\beta_1}-\frac{1}{\beta_2} - \frac{1}{\beta_3}\right) \left(\tanh \left(g(0) \right) - \frac{1}{3} \tanh ^3 \left(g(0)\right) \right)}.
\end{align}

\begin{figure}[h!]
\centering
\includegraphics[width=11cm]{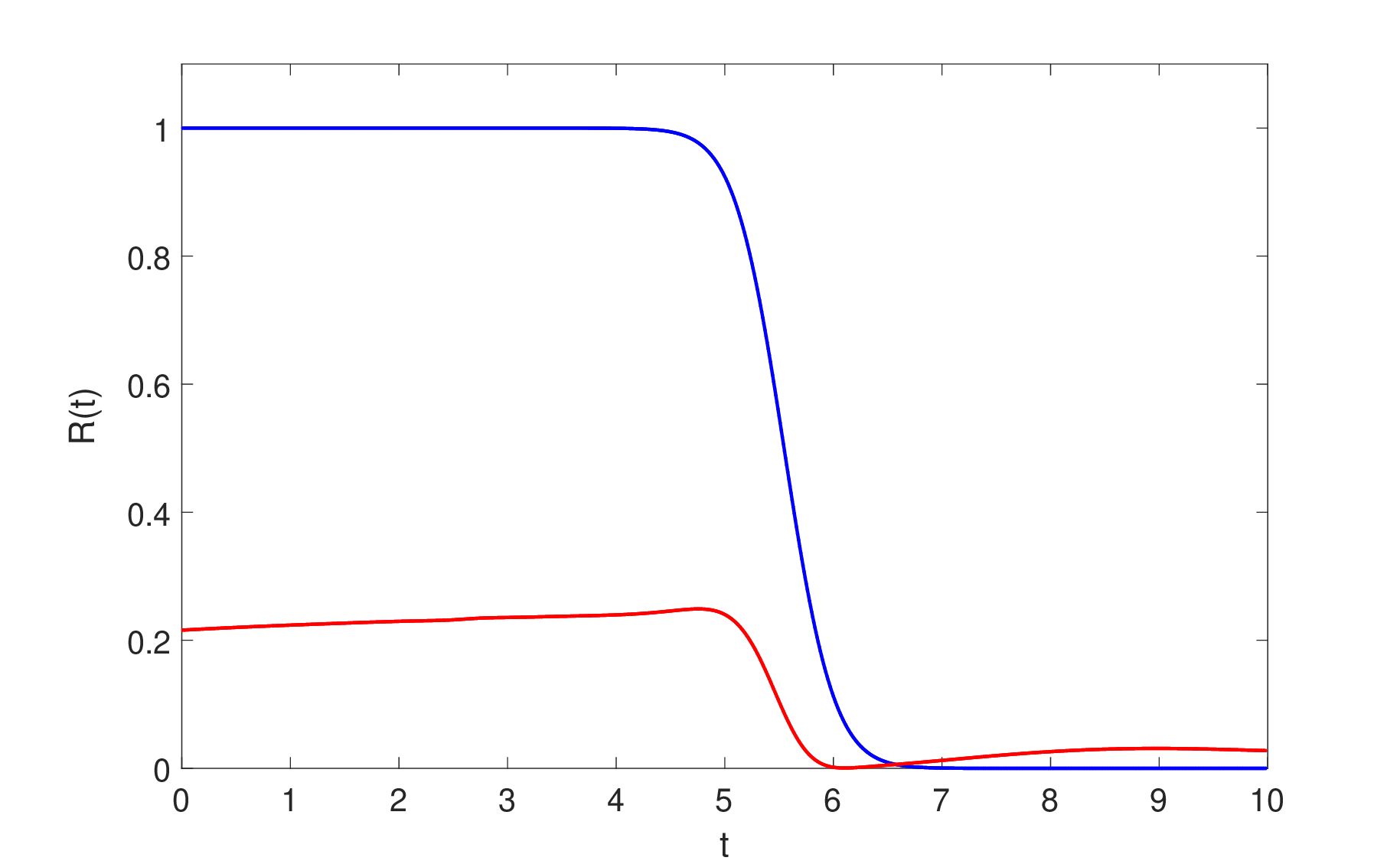}
\caption{(Color online) Time dependence of the reflection coefficient for the cases when the sum rule is satisfied (blue line) and broken (red line).}
\label{pic5}
\end{figure}

In Fig.~\ref{pic5} the dependence of $R$ on $t$ is plotted as a function of $t$ for the cases when the sum rule is satisfied with $\beta_1=1.2,\,\beta_2=2,\,\beta_3=3$ and broken with $\beta_1=0.5,\,\beta_2=2,\,\beta_3=3$. This systematic analysis clearly demonstrates that the absence of reflection occurs only when the sum rule is satisfied.

%%%%%%%%%%%%%%%%%%%%%%%%%%%%%%%
\section{Extensions to other graph topologies}

\begin{figure}[t!]
\centering
\includegraphics[width=10cm]{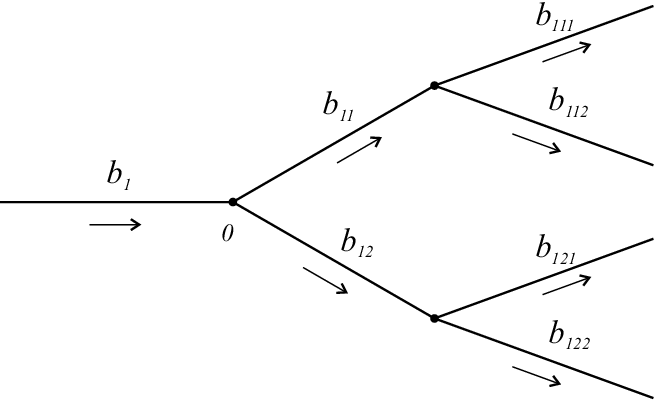}
\caption{A tree graph.}
\label{pic6}
\end{figure}

In the preceding sections, our analysis was restricted to the star graph, which can be regarded as a fundamental building block for more intricate graph topologies. Nevertheless, the proposed approach can be naturally extended to more complex configurations. To demonstrate this, we consider its application to a tree graph, as illustrated in Fig.~\ref{pic6}. 

This tree graph can formally be viewed as comprising three hierarchical branching levels, consisting of the bonds  
$b_1 \sim(-\infty;0],\, b_{1i}\sim[0;L_i]$ and
$b_{1ij}\sim[0;+\infty)$, where $i,j=1,2$. The nonlinear Klein-Gordon equation is given on the each bond $e\in\{1, 1i, 1ij\}$ of the graph and is written as
\begin{equation}
\partial_t^2 u_e - \partial_x^2 u_e - u_e + \beta_e u_e^3=0. \label{kgeq2}
\end{equation}
The following boundary conditions are imposed at the vertices:
\begin{eqnarray}
\left\{%
\begin{split}
 &\alpha_1 u_1 \big|_{x=0}= \alpha_{11} u_{11} \big|_{x=0} = \alpha_{12} u_{12} \big|_{x=0} ; \\
 &\frac{1}{\alpha_1} \partial_x u_1 \big|_{x=0} = \frac{1}{\alpha_{11}} \partial_x u_{11} \big|_{x=0} + \frac{1}{\alpha_{12}} \partial_x u_{12} \big|_{x=0}. \\
 &\alpha_{11} u_{11} \big|_{x=L_1} = \alpha_{111} u_{111} \big|_{x=0} = \alpha_{112} u_{112} \big|_{x=0} ; \\
 &\frac{1}{\alpha_{11}} \partial_x u_{11} \big|_{x=L_1} = \frac{1}{\alpha_{111}} \partial_x u_{111} \big|_{x=0} + \frac{1}{\alpha_{112}} \partial_x u_{112} \big|_{x=0}. \\
 &\alpha_{12} u_{12} \big|_{x=L_2}= \alpha_{121} u_{121} \big|_{x=0} = \alpha_{122} u_{122} \big|_{x=0} ; \\
 &\frac{1}{\alpha_{12}} \partial_x u_{12} \big|_{x=L_2} = \frac{1}{\alpha_{121}} \partial_x u_{121} \big|_{x=0} + \frac{1}{\alpha_{122}} \partial_x u_{122} \big|_{x=0}.
\end{split}%
\right.\label{bc1}
\end{eqnarray}

The kink (antikink) soliton solution of equation (\ref{kgeq2}) can
be written as
\begin{equation}
u_e (x,t) =\mp \frac{1}{\sqrt{\beta_e}} \, {\rm tanh} \left( \frac{x -
x_{0,e} - \upsilon t}{\sqrt{2(1 - \upsilon^2) }}
\right)\label{sol2}.
\end{equation}
By fixing the initial position of the kink (antikink) center at $l$ one can define $x_{0,e}$ as $x_{0,1} = x_{0,11} = x_{0,12} = l,\,x_{0,111} = x_{0,112} = l -
L_1,\,x_{0,121} = x_{0,122} = l - L_2$. Then, by requiring that the boundary conditions \eqref{bc1} are satisfied, we obtain the following sum rules for the nonlinearity coefficients
\begin{align}
\begin{split}
\frac {1}{\beta_1} &= \frac {1}{\beta_{11}} + \frac {1}{\beta_{12}},\\
\frac {1}{\beta_{11}} &= \frac {1}{\beta_{111}} + \frac {1}{\beta_{112}},\\
\frac {1}{\beta_{12}} &= \frac {1}{\beta_{121}} + \frac {1}{\beta_{122}}.
\end{split}
\label{sr-tree}
\end{align}

\begin{figure}[t!]
\centering
\includegraphics[width=12cm]{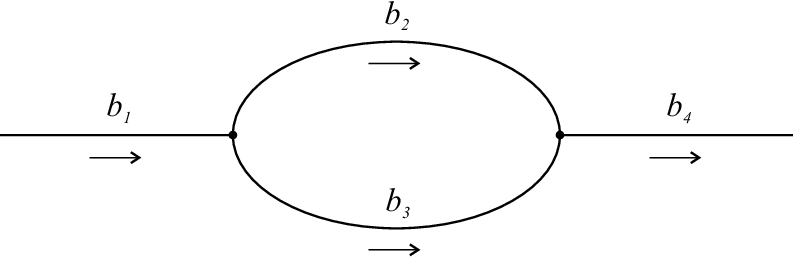}
\caption{A loop graph}
\label{pic7}
\end{figure}

Another topology, which is considered in this section is the loop graph plotted in Fig.~\ref{pic6}. It is composed of four bonds: $b_1\sim(- \infty ; 0] , \, b_{2}, b_3 \sim [0; L] $ and $ b_4 \sim [L; + \infty)$. On each bond of the loop graph
the nonlinear Klein-Gordon equation can be written by (\ref{kgeq2}) together with the boundary conditions at the vertices given as
\begin{equation}
\left\{%
\begin{split}
 &\alpha_1 u_1 \big|_{x=0}= \alpha_{2} u_{2} \big|_{x=0} = \alpha_{3} u_{3} \big|_{x=0} ; \\
 &\frac{1}{\alpha_1} \partial_x u_1 \big|_{x=0} = \frac{1}{\alpha_{2}} \partial_x u_{2} \big|_{x=0} + \frac{1}{\alpha_{3}} \partial_x u_{3} \big|_{x=0}. \\
 &\alpha_{2} u_{2} \big|_{x=L} = \alpha_{3} u_{3} \big|_{x=L} = \alpha_{4} u_{4} \big|_{x=0} ; \\
 &\frac{1}{\alpha_{2}} \partial_x u_{2} \big|_{x=L} + \frac{1}{\alpha_{3}} \partial_x u_{3} \big|_{x=L} = \frac{1}{\alpha_{4}} \partial_x u_{4} \big|_{x=0}. \label{bc2}
\end{split}%
\right.
\end{equation}

The kink (antikink) soliton solution is given by Eq.~(\ref{sol2}), where $x_{0,1} = x_{0,2} = x_{0,3} = l,\,x_{0,4} = l-L$. Again, imposing the boundary conditions (\ref{bc2}) one obtains the sum rule as
\begin{equation}\label{sr-loop}
\frac {1}{\beta_1} = \frac {1}{\beta_{2}} + \frac
{1}{\beta_{3}}=\frac {1}{\beta_{4}}.
\end{equation}

Fulfilling the sum rules for the nonlinearity coefficients \eqref{sr-tree} and \eqref{sr-loop} ensures that Eq.~\eqref{sol2} is an exact solution of the NLKG equation \eqref{kgeq2}, subject to the boundary conditions \eqref{bc1} and \eqref{bc2} imposed for the tree and loop graphs, respectively. Moreover, under these constraints, the propagation of the kink (or antikink) in the direction of increasing coordinates occurs without reflection at the graph vertices.

\section{Summary}

In this paper, we studied the nonlinear Klein–Gordon equation on metric graphs, focusing primarily on the star graph with three semi-infinite bonds. By imposing generalized continuity and current conservation conditions at the branching point (vertex), we derived a constraint for the nonlinearity coefficients in the form of a sum rule that ensures conservation of total energy.

We began by deriving nonlinear boundary conditions directly from the energy conservation law. These boundary conditions yield weighted continuity of the wave function and a Kirchhoff-type condition on its derivatives at the vertex. Under these conditions, we obtained an exact soliton solution for the NLKG equation on the star graph and identified a constraint — in the form of the sum of inverse nonlinearity coefficients — that ensures reflectionless transmission through the vertex. We derived these conditions for a star graph with three bonds, and the extension to the general case with $N$ bonds follows directly.

Using this soliton solution, we further demonstrated that the total momentum of the system is conserved. The conservation of total energy was confirmed analytically, and the time dependence of the reflection coefficient was computed. Our analysis shows that reflection at the branching point is absent only when the sum rule is satisfied.

Finally, we extended our approach to more complex graph structures, including tree and loop graphs, formulating corresponding boundary conditions and sum rules for the nonlinearity coefficients that ensure energy conservation and reflectionless soliton propagation.

\section*{Appendix}
\subsection*{A. Derivation of nonlinear boundary conditions from linear ones}
In this appendix we show that the nonlinear boundary conditions in Eq.~\eqref{nbc1} can be derived from the two linear boundary conditions  \eqref{wc1}-\eqref{kr1}. By differentiating Eq.~\eqref{wc1} with respect to time, one obtains 
$$
 \alpha_1 \partial_t u_1 \big|_{x=0} = \alpha_2 \partial_t u_2 \big|_{x=0} = \alpha_3 \partial_t u_3 \big|_{x=0}.
$$
Multiplying Eq.~\eqref{kr1} to $\alpha_1 \partial_t u_1 \big|_{x=0}$ results in
$$
 \frac{1}{\alpha_1} \partial_x u_1 \big|_{x=0} \alpha_1 \partial_t u_1 \big|_{x=0} = \frac{1}{\alpha_2} \partial_x u_2 \big|_{x=0} \alpha_1 \partial_t u_1 \big|_{x=0} + \frac{1}{\alpha_3} \partial_x u_3 \big|_{x=0} \alpha_1 \partial_t u_1 \big|_{x=0}.
$$

Using differentiated weighted continuity condition above, one obtains modified weighted current conservation condition: 
$$
\frac{1}{\alpha_1}\partial_x u_1 \alpha_1 \partial_t u_1 |_{x=0} = \frac{1}{\alpha_2} \partial_x u_2 |_{x=0} \alpha_2 \partial_t u_2 |_{x=0} + \frac{1}{\alpha_3} \partial_x u_3 |_{x=0} \alpha_3 \partial_t u_3 |_{x=0},
$$
which results in the nonlinear boundary condition
$$
\partial_x u_1 \partial_t u_1 |_{x=0} = \partial_x u_2 \partial_t u_2 |_{x=0} + \partial_x u_3 \partial_t u_3|_{x=0}.
$$

\subsection*{B. The numerical scheme for solving the NLKG equation on the metric star graph with three semi-infinite bonds}

Based on the numerical scheme proposed in \cite{Ablowitz3}, we use the following notations for the restriction of $u_j(x,t)$ to a square mesh lattice:
\begin{eqnarray*}
    &&u_{j,n}^m=u_j(nh,mh),\label{neq1}\\
    &&v_{j,n}^m=u_j\left((n+\frac{1}{2})h,(m+\frac{1}{2})h\right),\\
    &&w_{j,n}=\partial_tu_j(nh,0).\label{neq3}
\end{eqnarray*}
Then from the Taylor series we have
\begin{eqnarray*}
    v_{j,n}^0=\frac{1}{2}(u_{j,n}^0+u_{j,n+1}^0)+\frac{h}{4}(w_{j,n}+w_{j,n+1})-\frac{h^2}{8}F'\left(\frac{u_{j,n}^0+u_{j,n+1}^0}{2}\right)+O(h^3),\,n=\overline{0,N-1}\label{neq4}
\end{eqnarray*}
where $F(g)=-\frac{1}{2}g^2+\frac{\beta_j}{4}g^4$, and subsequently we can find
\begin{eqnarray*}
    u_{j,n}^{m+1}=-u_{j,n}^m+v_{j,n}^m+v_{j,n-1}^m-\frac{h^2}{4}F'\left(\frac{v_{j,n}^m+v_{j,n-1}^m}{2}\right)+O(h^4),\,n=\overline{1,N-1}.\label{neq5}
\end{eqnarray*}
Discretization of the boundary conditions (\ref{wc1}) and (\ref{kr1}) results in the following matrix relation
\begin{eqnarray*}
    \left(\begin{array}{ccc}u_{1,N}^{m+1}\\u_{2,0}^{m+1}\\u_{3,0}^{m+1}\end{array}\right)=A^{-1}B,
\end{eqnarray*}
where $A=\left(\begin{array}{ccc}\alpha_1&-\alpha_2&0\\\alpha_1&0&-\alpha_3\\\frac{1}{\alpha_1}&\frac{1}{\alpha_2}&\frac{1}{\alpha_3}\end{array}\right)$, $B=\left(\begin{array}{ccc}0\\0\\\frac{1}{\alpha_1}u_{1,N-1}^{m+1}+\frac{1}{\alpha_2}u_{2,1}^{m+1}+\frac{1}{\alpha_3}u_{3,1}^{m+1}\end{array}\right)$. 

Finally, we have
\begin{eqnarray*}
    v_{j,n}^{m+1}=-v_{j,n}^m+u_{j,n+1}^{m+1}+u_{j,n}^{m+1}-\frac{h^2}{4}F'\left(\frac{u_{j,n+1}^{m+1}+u_{j,n}^{m+1}}{2}\right)+O(h^4),\,n=\overline{0,N-1}.\label{neq6}
\end{eqnarray*}

%%%%%%%%%%%%%%%%%%%%%%%%%%%%%%%%%%%%%%%%%%%%%%%%%
\section*{Acknowledgements}
The work of JY is supported by the grant of the Innovation Development Agency of the Republic of Uzbekistan (Ref. No. F-2021-440).


\begin{thebibliography}{99}


\bibitem{Ablowitz1}  %1
Ablowitz M.J. and Segur H. Solitons and the Inverse Scattering Transform. SIAM, Philadelphia, 1981, 425 p. 

\bibitem{Ablowitz2}  %2
Ablowitz M.J. and Clarkson P.A. Solitons, Nonlinear Evolution Equations and Inverse Scattering.  Cambridge University Press, Cambridge, 1991, 513 p.

\bibitem{Greiner} %3
Greiner W. Relativistic Quantum Mechanics-Wave Equations. 3rd edition. Springer-Verlag, Berlin, 2000, 345 p.

\bibitem{Grundland} %3
Grundland A.~M. and Infeld E. A family of nonlinear Klein–Gordon equations and their solutions. J. Math. Phys. 33, 2498–2503 (1992)

\bibitem{Matsuno} %3
Matsuno Y. Exact solutions for the nonlinear Klein–Gordon and Liouville
equations in four‐dimensional Euclidean space. J. Math. Phys. 28, 2317–2322 (1987)

\bibitem{Clarkson} %3
Clarkson, P.~A., McLeod, J.~B., Olver, P.~J., \& Ramani, A. Integrability of Klein–Gordon Equations. SIAM Journal on Mathematical Analysis, 17(4), 798–802 (1986).

\bibitem{Ablowitz3}  %4
Ablowitz M.J., Kruskal M.D. and Ladik J.F. Solitary wave collisions. SIAM J. Appl. Math., 1979, 36 (3), 428-437.

\bibitem{Jim} %5
Jim\'enez S.,V\'azquez L. Analysis of Four Numerical Schemes for a Nonlinear Klein-Gordon Equation. App. Math. and Comp., 1990, 35, 61-94.

\bibitem{Los}    %6
Los Vu-Quoc and Shaofan Li. Invariant-conserving finite difference algorithms for the nonlinear Klein-Gordon equation. Computer Meth. in App. Mech. and Eng., 1993, 107(3), 341-391.

\bibitem{Sobirov1} Z.Sobirov, D.Matrasulov, K.Sabirov, S.Sawada, and K.Nakamura. Integrable nonlinear Schr\"odinger equation on simple networks: Connection formula at vertices. Phys. Rev. E, 2010, 81, 066602.

\bibitem{Adami}    %8
Adami R., Cacciapuoti C., Finco D. and Noja D. Fast solitons on star graphs. Reviews in Math. Phys., 2011, 23(4), 409-451.

\bibitem{Sobirov2} Z.Sobirov, D.Babajanov, D.Matrasulov, K.Nakamura, and H.Uecker. Sine-Gordon solitons in networks: Scattering and transmission at vertices. EPL, 2016, 115, 50002.

\bibitem{Sabirov1}   %10
Sabirov K.K., Babajanov D.B., Matrasulov D.U. and Kevrekidis P.G. Dynamics of dirac solitons in networks. J. Phys. A: Math. Theor., 2018, 51, 435203.

\bibitem{Sabirov2}  %11
Sabirov K.K., Yusupov J.R., Matyokubov Kh.Sh., Susanto H., Matrasulov D.U. Networks with point-like nonlinearities. Nanosystems: Phys. Chem. Math., 2022, 13(1), 30-35.

\bibitem{Aripov2019}  
Aripov M.M., Sabirov K.K., Yusupov J.R., Transparent vertex boundary conditions for quantum graphs: simplified approach. Nanosystems: Phys. Chem. Math., 2019, 10(5), 505-510.

\bibitem{Yusupov2019}  
Yusupov J.R., Sabirov K.K., Ehrhardt M., Matrasulov D.U., Transparent nonlinear networks. Phys. Rev. E, 2019, 100(3), 032204.

\bibitem{Kottos} T.Kottos and U.Smilansky. Periodic Orbit Theory and Spectral Statistics for Quantum Graphs. Annals of Physics, 1999, 274(1), 76-124.

\bibitem{Gnutzmann} S.Gnutzmann and U.Smilansky. Quantum graphs: Applications to quantum chaos and universal spectral statistics. Advances in Physics, 2006, 55, 527-625.

\bibitem{Berkolaiko} %3
Berkolaiko G. and Kuchment P. Introduction to quantum graphs. (No. 186). American Mathematical Soc. 2013, 275 p.

\bibitem{Maier}  %14
Maier D. Construction of breather solutions for nonlinear Klein-Gordon equations on periodic metric graphs. J. Differential equation, 2019, 268(6). 2491-2509.

\bibitem{Kumbi1}  %15
Kumbinarasaiah S., Ramane H.S., Pise K.S., Hariharan G. Numerical Solution for Nonlinear Klein–Gordon Equation via Operational-Matrix by Clique Polynomial of Complete Graphs. Int. J. Appl. Comput. Math., 2021, 7(12), 1-19.

\bibitem{Takei}  %16
Takei Y., Iwata Y. Space-time breather solution for nonlinear Klein-Gordon equations. J. Phys.: Conf. Ser., 2021, 1730, 012058.

\bibitem{Golosh1}  %17
Goloshchapova N. A nonlinear Klein–Gordon equation on a star graph. Mathematische Nachrichten, 2021, 294, 1–23.

\bibitem{Alrazi}  %18
Alrazi A., Harun-Or R. and Abdullah A. Bright, Dark, and Rogue Wave Soliton Solutions of the Quadratic Nonlinear Klein–Gordon Equation. J. Symmetry, 2022, 14, 1223.

\bibitem{strauss1989} 
Strauss, W. A. (1989) Nonlinear wave equations, CBMS regional conference series in mathematics, vol 73. American Mathematical Society, Providence.

\bibitem{Wazwaz} 
Wazwaz A.-M. Compactons, solitons and periodic solutions for some forms
of nonlinear Klein–Gordon equations. Chaos, Solitons and Fractals 28 (2006) 1005–1013.

\bibitem{Bratsos} 
Bratsos, A.G. On the numerical solution of the Klein-Gordon equation. Numer. Methods Partial Differential Eq., 2009, 25: 939-951.







\end{thebibliography}
\end{document}